# A polymorphous band structure model of gapping in the antiferromagnetic and paramagnetic phases of the Mott insulators MnO, FeO, CoO, and NiO

Giancarlo Trimarchi [1] and Alex Zunger [2]

(1) Department of Chemistry, Northwestern University, Evanston, Illinois 60208, USA

(2) Renewable and Sustainable Energy Institute, University of Colorado, Boulder, Colorado 80309, USA

A band structure description of the observed large band gaps and moments in both the antiferromagnetic (AFM) and paramagnetic (PM) phases of the classic NaCl-structure Mott insulators MnO, FeO, CoO, and NiO is provided by ordinary, single-determinant density functional theory (DFT) method. This is enabled by permitting unit cells that can lift the degeneracy of the $d$ orbitals and develop large on-site magnetic moments without violating the global, averaged NaCl symmetry. As noted by previous authors, the ordered AFM phases already show in band theory significant band gaps when one uses the observed NaCl crystal structure but doubles the unit cell by permitting different potentials for transition metal atoms with different spins; for the degenerate $d$ band cases of CoO and FeO, energy-lowering atomic displacements remove the band degeneracies, whereas for MnO and NiO the spin-dependent crystal field symmetry already does so. However, for the disordered PM phases the commonly used band model has been to assume the macroscopically observed, averaged NaCl structure, where all transition metal (TM) sites are forced to be symmetry-equivalent (a *monomorphous* description); for the PM phase this forces zero moment *on an atom by atom basis*, thus producing a gapless PM state, in sharp conflict with experiment. We do not follow this description. Instead, we allow larger NaCl-type supercells where each TM site can have different local bonding and spin environments (a *polymorphous* description) and thus the geometric flexibility to acquire symmetry-lowering distortions that lower the total energy and can break the symmetry of the $d$ orbitals. It turns out that such a polymorphous description of the structure (the existence of a distribution of different local spin and bonding environments) allows large on-site magnetic moments to develop spontaneously in the self-consistent DFT+U leading to significant (1-3 eV) band gaps in the AFM, FM, and PM phases of the classic NaCl-structure Mott insulators MnO, FeO, CoO, and NiO. We adapt to the spin disordered configurations in the PM phases the "special quasi-random structure" (SQS) construct known from the theory of random substitutional alloys whereby supercell approximants which represent the best random configuration average (not individual snapshots) for finite (64, 216 atoms or larger) supercells of a given lattice symmetry are constructed. Thus, avoiding a monomorphous description of the disordered magnetic phases allows even ordinary DFT+U, which represents the $N$ electron system with a single-determinant wavefunction, to describe the gapping not only in the AFM phases but also in the PM phases of the classic Mott insulators MnO, FeO, CoO, and NiO.

Email: gtrimarc@gmail.com, alex.zunger@colorado.edu

# I. Introduction

The physical origin of the insulating phases occurring in crystals with partially occupied $d$ shells exemplified by the transition metal (TM) monoxides MnO, FeO, CoO and NiO, has held the condensed matter physics community in constant fascination ever since Mott and Peierls proposed an explanation[1]. These oxides have a low-temperature antiferromagnetic (AFM) phase, in which they exhibit slightly distorted rock salt structures, and a high-temperature paramagnetic (PM) phase, having macroscopically the cubic rock salt structure and a globally zero magnetic moment. In simplified band structure calculations[2-7] it has been customary to evaluate the electronic structure for the macroscopically observed average rock salt configuration $\mathbf{\Sigma}_0$, leading to zero local magnetic moment $\mu_i(\mathbf{\Sigma}_0)$ at *each metal site i*, and therefore, by symmetry, to zero band gaps $E_g(\mathbf{\Sigma}_0)$. Here, $\mathbf{\Sigma}_0$ was taken as the non-magnetic, cubic rock salt configuration in which all TM sites are equivalent (a *monomorphous* representation). As is generally taught[8, 9], the ensuing electronic structure of compounds having partially occupied energy bands described in a structure where all atoms are equivalent would be metallic[2, 3] with the Fermi level intersecting a band. Yet, experimentally MnO, FeO, CoO and NiO are local-moment large band gap insulators, both in the AFM and PM phase[10-12] (see Table I).

The fundamental disagreement between such band structure theory and experiment set the historical stage for modeling the electronic structure of the PM phases of MnO, FeO, CoO and NiO by many-body, correlated electron descriptions, such as the description based on the Hubbard Hamiltonian[13, 14], or, more recently, dynamical mean field theory (DMFT)[15, 16]. Within such theories, the gap of the AFM and PM phases of these oxides emerges because the $d$ electrons become localized due to the correlation-induced electron-electron repulsion, even in the absence of spatial symmetry breaking (symmetry can break afterwards, as a secondary fact). From the strongly correlated standpoint the existence of local magnetic moments is a consequence of the electron localization and not an essential part of the gap opening mechanism itself. The computational machinery required to model the band gap opening in the PM phases of such Mott insulators is formulated to treat dynamic correlation, and must therefore be based on a many-body treatment of the electronic structure that goes fully beyond Bloch-periodic, single-Slater determinant band theory. Indeed, DMFT descriptions[4, 16] of correlated phenomena often motivate the need for such techniques by demonstrating qualitative failures of given band theories.

In the density functional formalism (DFT) the *exact ground-state energy* and spin densities of the interacting electrons in an external spin-dependent potential $v_\sigma(\mathbf{r})$ can be found (even for correlated systems) from an effective one-electron Schrödinger equation in a single-determinant approach *if* the exact spin-density functional $E_{xc}[n^\uparrow(\mathbf{r}), n^\downarrow(\mathbf{r})]$ were given. While the exact functional is presently unknown, it would seem worthwhile to examine whether the Kohn-Sham DFT, with the current approximate exchange and correlation functional, could capture the central phenomenology of the appearance of band gaps and local moment in the PM phases for the classic Mott insulators MnO, FeO, CoO and NiO. The answer to this question has been previously clouded by the fact that often the structural input to such band structure calculations was restricted to the macroscopically averaged configuration (referred to here as the 'monomorphous representation', in the present case the NaCl structure), where each metal site sees an identical local environments and potentials. Indeed, because a monomorphous representation of a disordered PM phase forces upon us in band theory a zero magnetic moment on an atom-by-atom basis, the ensuing band gaps were always zero, irrespective of the quality of the description of the interelectronic interaction $E_{xc}[n^\uparrow(\mathbf{r}), n^\downarrow(\mathbf{r})]$ in KS-DFT band theory. We show here that symmetry breaking afforded by using sufficiently flexible unit cells (the polymorphous representation where chemically identical metal atoms are allowed to have different local environments) produces in DFT+$U$ [17-21] an insulating solution with strong local magnetic moments, in agreement with experiment. The conventional physical chemistry encoded in DFT with current functionals-- bonding, magnetism, spin-polarization, Jahn-Teller distortions, hybridization of the components of $e_g$ and $t_2$ orbitals --suffices to produce large moments if not disallowed by symmetry. We believe that the fact that the main attributes of the classic Mott insulators (i.e., the existence of gaps) could be described by *mean-field, single configuration band theory* is highly consequential, as it redefines the minimal level of computational effort required.

An often used, pragmatic definition of correlation is "everything that DFT does not get right". Given the theorem recalled above, the domain so defined is a shrinking proposition. Indeed, the ability of generalized KS-DFT to reproduce the insulating character of the paramagnetic phases of MnO, FeO, CoO, and NiO when applied using a polymorphous structural representation, does not imply that these are not correlated materials. We further note that it is entirely possible that some other Mott compounds would remain metallic in a single-particle treatment even if the monomorphous description were replaced by a polymorphous representation. It is also possible that KS-DFT with the current functional could miss properties other than the insulating character of the PM phases. We are not claiming otherwise. These are open, future research questions.

## II. Dual inputs to electronic structure theory

Any electronic structure method requires specifying (a) a representation for the *crystal structure* (and, for random systems such as the PM phases, the way the configurational average is performed), as well as (b) the *type of electronic interactions allowed* by the Hamiltonian and its solver (e.g., forms of exchange and correlation in band theory; dynamic correlation in explicitly correlated theories). In regard to (b), Table I shows the magnetic moments and band gaps calculated by ordinary DFT+$U$ for the AFM as well as the assumed ferromagnetic (FM) phases of MnO, FeO, CoO, and NiO. Recall that in the DFT+$U$ method[21] as well as related methods such as the self-interaction corrected DFT[22] and the hybrid functionals[23] are all single particle schemes in which the wave function of the $N$ electron system is a single determinant. *We see from Table I that even in the single-determinant DFT+U description with reasonable U=5 eV, which we used across the board for all compounds and spin configurations (we make no attempt at optimizing or fitting), these transition-metal monoxides, regardless of the type of magnetic ordering (AFM or FM), exhibit large local moments and band gaps.* This opens the possibility that the actual magnetic order may not be the primary reason for these materials to be insulators, but the existence of on-site magnetic moments may.

## III. Allowing for a polymorphous description of the magnetic structure of the paramagnetic phases as a possible route to obtaining the insulating character

Here, we wish to examine if significant on-site magnetic moments might produce a gap also in the PM phases within a single-determinant approach, had these moments not been eliminated at the outset by selecting for the representation for the crystal structure [(a) above] the monomorphous macroscopically averaged configuration $\Sigma_0$ where each TM site sees the same local environment. In the latter case the global zero moment characteristic of the PM phase is interpreted on an atom by atom basis so $\mu_i(\Sigma_0) = 0$ at each metal site $i$, leading in band theory (where large moments mean large exchange splitting) to a band gap $E_g(\Sigma_0)=0$. The polymorphous approach allows each TM atom to see a distinct local magnetic environment and a locally varying density-functional potential $V[\rho(r), m(r)]$ [where $\rho(r)$ and $m(r)$ are, respectively, the electron density and magnetization at position $r$] subject to the constraint that the *total* magnetic moment is zero as must be in a paramagnet. We will enquire if such a representation has sufficient geometrical freedom (e.g., unrestricted spatial symmetry) to allow in a self-consistent DFT(+U) calculation the evolution of local magnetic moments on individual sites, if this would lower the total energy. In band theory language, large moments imply large exchange splitting which enable large gaps.

The possibility of gapping a metal by permitting the geometric flexibility that enables electronic symmetry lowering is familiar in other cases. For example, the DFT total energy of cubic perovskite $BaBiO_3$ with a single Bi site per cell (a metal), is lowered by doubling the cubic perovskite primitive cell, allowing two $Bi^{4+}$ ions to express their multivalent nature by disproportioning into $Bi^{3+}$ + $Bi^{5+}$ (see Ref.[24]) with each site having its own, local bond geometry, a cell-internal symmetry lowering that caused gapping. Similarly (see Ref.[25]) for $CsTlF_3$ being allowed to express the multivalence of Tl in $Cs_2$ $[Tl^{1+}Tl^{3+}]F_6$ leading to gapping of a previously metallic state. In all such cases, a restricted structural description (one type of octahedron) incorrectly produced a metal, whereas a more flexible description of the cell lowered the total energy and produced the observed insulating gap in band theory. The paramagnetic phase of the TM monoxides also exhibits gapping when the restrictions on the structural description leading to metallic states are relaxed.

### A. Using the average $\langle P \rangle$ of the properties $\{p_i\}$ of individual configurations $\{\sigma_i\}$ vs. using the property $P(\Sigma_0)$ of the average configuration $\Sigma_0 = \langle \sigma_i \rangle$

A common approximation in calculating observable macroscopic electronic or magnetic property $\langle P \rangle$ of a chemically disordered random alloy $A_x B_{1-x}$ of composition $x$ or of a disordered moment PM phase is to substitute the calculation of the ensemble average $\langle P \rangle$ for property $P$ with the calculation of the property $P(\boldsymbol{\Sigma_0})$ of the macroscopically averaged configuration $\boldsymbol{\Sigma_o}$. This monomorphous approximation has been used in the single-site Coherent Potential Approximation (s-CPA) for chemical alloys[26] where all A (B) sites in the random alloy see the same potential $V_A$ ($V_B$), irrespective of the existence of different local environments for different A sites (characterized by different numbers of A vs. B nearest neighbors). This approach in alloy theory forced vanishing charge fluctuations (hence zero Madelung contribution to the total energy[27, 28]) and vanishing atomic displacements,[29, 30, 31, 32] both in conflict with more general theories (such as supercells[33],[27, 29, 34] having a polymorphous distribution of different A sites each with its local environment (and same for B sites)

The correct way to calculate the property $P$ of a phase that can have numerous individual configurations $\{\boldsymbol{\sigma}_i^{(n)}\}$ each with property $P(\boldsymbol{\sigma}_i^{(n)})$, (where $n$ is the index of the configuration), is to calculate the polymorphous statistical average $\langle P \rangle = \sum_n c_n P(\boldsymbol{\sigma}_i^{(n)})$ over the ensemble of microscopic spin configurations accessible to the system, instead of assuming $\langle P \rangle = P(\boldsymbol{\Sigma_o})$. The former approach has largely replaced the monomorphous approaches (s-CPA, Virtual Crystal Approximation) to the theory of disordered substitutional alloys $A_xB_{1-x}$, producing qualitatively different results[27, 29, 33, 34] in substantial agreement with experiment, when available. The same polymorphous approach can be applied to spin disordered phases. In a spin-disordered phase the *orientation* of the on-site magnetic moments $\boldsymbol{\mu}_i$ can change over time showing spin wave excitations and decides the low energy scale of the problem; such fluctuations in the orientation of the moments have zero overall average. This is confirmed by neutron diffraction measurements that show that the PM phases exhibit a spectrum of sharp Bragg peaks corresponding to the ideal rock salt structure. However, the *magnitude* of the on-site moments $|\boldsymbol{\mu}_i|^2$ in gapped systems will not be zero, deciding the higher energy scale of the problem. As a result, the time average of the gaps of all configurations could be different than zero. Here, we substitute the time averages with ensemble averages, (a step that is conceptually analogous to that underpinning the formulation of the Disordered Local Moment (DLM) method). We will thus examine whether and how accurately the gap of the PM phases of MnO, FeO, CoO, and NiO can be predicted within a single-determinant description, for which here we use the DFT+U scheme, if one correctly estimates the statistical average $\langle P \rangle = \sum_n c_n P(\boldsymbol{\sigma}_i^{(n)})$ over the ensemble of microscopic spin configurations accessible to the system instead of forcing a zero moment on an atom-by-atom basis.

### B. The special quasi-random structure (SQS) as a finite supercell realization of a polymorphous paramagnet, not a snapshot configuration

Let us focus on the band gap as the property $P$ to be calculated. Instead of calculating the band gap for many snapshot configurations $\{\boldsymbol{\sigma}_i^{(n)}\}$ and averaging the corresponding band gaps $P(\boldsymbol{\sigma}_i^{(n)})$, we construct a single supercell of $N$ sites that *approximates the polymorphous configurational average*. This is done by requiring that the pair and multibody atom-atom correlation functions in this special $N$ site cell best match the analytically known correlation functions for the infinite, perfectly random configuration.[35, 36] Convergence with respect to $N$ must be examined; we use $N \leq 216$ atoms/cell finding that the moments and the total energy have stabilized. The SQS fully complies with the polymorphous description of the PM phases that we want to apply here. An observable $P$ calculated for such a structure is not simply the property of a single snapshot configuration but approximates the ensemble average $\langle P \rangle$ for the random configuration (see Appendix A for the SQS construct[37-39] and the explanation of how an SQS approximates the ensemble average for a random system).

It is clear that describing random alloys by periodic structures will introduce spurious correlations beyond a certain distance ("periodicity errors"). However, many physical properties of solids are characterized by microscopic length scales that can be ordered according to their typical size so as to establish a hierarchy. For instance, interactions between distant neighbors generally contribute less to the total energy than do interactions between close neighbors. Therefore, the guiding idea in the construction of special quasi-random structures is to obtain within such structures a close reproduction of the perfectly random network for the first few shells around a given site, while the periodicity errors originate from the arrangement of the more distant neighbors. In this respect, the SQS construct is reminiscent of the 'special **k** points' used for Brillouin zone integration[40, 41] in the sense that the selected **k** points are not meant to reproduce properties that reflect mostly the long-range order. The accuracy of the SQS improves as one uses larger SQS cell representation (analogous to using more k points in BZ sampling methods) in which longer range correlation functions can be matched.

The SQS, as we just pointed out, is a convenient computational tool to approximate ensemble averages. It has been shown that relatively small SQS produce numerically the same property values as far larger (ergodic) randomly selected supercells (see Ref.[42]). Note, however, that the SQS approach is not to be confused with the commonly practiced supercell approach. In the supercell approach, one occupies lattice sites by different spins using, say, a random statistics (i.e., via coin toss) or some choice of short-range order. However, each such occupation pattern corresponds to a single snapshot and in order to calculate the observable property $\langle P \rangle$, which is an ensemble average, one should average the properties $\{P_i\}$ of different supercell snapshots. In the SQS approach the property $P_\text{SQS}$ calculated for one SQS provides an approximation of the average $\langle P \rangle$ which is progressively improved by increasing the size $N$ of the SQS and by extending the order and size of the figures that the SQS algorithm tries to hierarchically match. Because the SQS is a polymorphous approach, it allows chemically identical sites to develop their own, energy-lowering displacement patterns. In the transition metal monoxides investigated here the minimization of the total energy for the PM phases shows negligible positional atomic displacements relative to the rock salt positions (less than 0.05 A in amplitude).

Fig.1 shows the SQS we use for the random PM phase. The histogram in Fig.1 illustrates that, while in the AFM phase (formed by doubling of the primitive rock salt cell) each metal atom has 6 spin-up and 6 spin-down metal neighbors [denoted by (6,6)], in the SQS representation of the high-temperature PM phase there is a *distribution of local environments*, e.g., (4,8), (6,6), (8,4) etc. The landscape of the self-consistent DFT potential $V[\rho(\boldsymbol{r}), m(\boldsymbol{r})]$ corresponding to the SQS, in effect, allows each metal site to experience its own distinct 'particle-in-a-box' type potential, simply because chemically identical metal sites that have different neighbors 'see' different local potentials. Just as the doubling of the primitive cell is needed to produce anti-ferromagnetism in the low temperature AFM phases, the magnetic SQSs allow to capture the different local patterns in the distribution of the magnetic moments that characterize the PM phases. In these magnetic SQSs the symmetry is fully removed on average at each site, enabling solutions in which the degeneracy of the *d* levels is completely lifted with the result of producing significant local moments. As we will see below, such polarization of the charge density into certain areas of space (not necessarily localization in the sense of 2 electrons on one site as in the Mott-Hubbard picture) is important in driving the selective occupation of certain *d* orbitals out of the originally degenerate ones. The consequence is the development of large energy gaps.

### C. The role of *U* in DFT+*U*

We used the PBE semi-local approximation to the exchange and correlation functional; for simplicity, we use a constant value of $U - J = 5.0$ eV (the parameter in the DFT+U formulation of Ref.[43]) for all materials in this study, although, most likely one can improve agreement with experiment by tweaking *U* separately for each compound. In the DFT+U method the DFT total-energy functional is corrected by two terms (Refs.[17-21, 43-47]). The first term is a mean-field approximation of the electron-electron interaction within a subset of localized orbitals (here, the d orbitals). The second term subtracts the contribution of the electron-electron interaction already accounted for in the approximate functional and largely consists of the self-Coulomb interaction (a manifested one body effect). The DFT+U method can be seen as a simplification of the self-interaction correction (SIC), which shifts down d states, while creating (just as the uncorrelated Hartree-Fock method) a distinction between occupied and unoccupied states, restoring the discontinuity in the total energy as a function of the number of electrons. This aspect is important for the description of open-shell systems treated in the current paper.

Despite the impression suggested by the letter (Hubbard) "*U*", the DFT+*U* method (perhaps better renamed DFT+*V* to avoid such a confusion) does not imply correlation *in the Hubbard Hamiltonian sense*. We note that the majority of practitioners apply these methods with the belief that they model many-body correlations. DFT+*U*, as well as the hybrid functionals and SIC DFT, are all methods in which the wavefunction of the N electron system is a single determinant. In single-determinant, band structure approaches each band structure calculation occupies its levels in a single specific manner by electrons (a single Slater determinant) and different possible patterns of occupation of levels by electrons (which can be built in separate band structure calculations) have no way of seeing each other.

### IV. Summary of the main computational results on the PM phases.

Before we discuss the physical picture that emerges we state the results obtained for the magnetically ordered AFM and FM phases and for the magnetically disordered PM phases, which we modeled by a polymorphous description. As we are not interested here in fitting the calculated gaps and moments to experiment, for simplicity we use DFT+$U$ with a generic constant $U$= 5 eV same for the FM, AFM, and PM phases and all compounds; we also assume collinear moments and neglect short-range order in the PM phase assuming perfect randomness (the high temperature limit). All such fine-tuning corrections can be used in the future if one seeks more accurate, material-dependent physics. The schematic of Fig. 2 summarizes the hierarchy of site-specific effects that remove the degeneracy in the four monoxides. The calculated values of moments and band gaps are listed in Table I and reported in the projected density of state plots of Figs. 3 and 4.

The key ingredient of the theory is allowing for chemically identical sites to develop their own unique local environments and potentials rather than forcing a monomorphous representation which leads in the PM phase to non-magnetic unit cells. (We note that the FM spin arrangement is monomorphous and has a gap due to its long-range order, but it's obviously not a good model for the paramagnetic phase that is magnetically disordered). The SQS construct allows chemically identical sites to develop their own unique local environments and potentials and is disordered, and hence serves as a good approximation for the configuration of the PM phase. We see that a straightforward band structure description with appropriate structural/configurational input and reasonable value of the self-Coulomb U parameter captures the moment formation and gapping in the AFM as well as PM phases of the classic Mott insulators.

## V. Analysis of the results

### A. Analysis of the occupations of the localized orbitals

For our analysis of the DFT+$U$ results we sought linear combinations of the $d$ orbitals that form a good representation of the point-group symmetry at the TM sites. The $d$-orbital occupation matrix that enters the "+$U$" term of the DFT+$U$ energy functional is calculated using the $t_{2g}$ and $e_g$ orbitals as basis. However, the actual magneto-crystalline order in the AFM phases, or the lack of it in the PM phases, breaks the cubic point-group symmetry at the TM sites. In such a case a good representation for the $d$ orbitals, which is often referred to as the "crystal field representation", is that defined by the eigenvectors of the occupations matrices. This representation is also meaningful in terms of the mechanism that drives the band gap opening. The sum of the probability distributions of the eigenvector functions with spin down each weighted by its occupation gives the distribution of the minority-spin electrons density around the transition metal sites, which we inspect in the following. See Appendix B.1 for more details on this representation.

### B. Making sure that the electronic structure DFT description does not get trapped in a high symmetry basin.

On the technical side, one needs to assure that the self-consistency cycle is conducted so as to avoid that it get trapped in a high-symmetry solution but to permit the exploration of broader positional as well as wave function symmetries. In the case of systems in which the crystal field produces degenerate states that are partially filled, one must explore lowered symmetries of the electronic state by allowing for distortions of the lattice.[20, 21, 48] Thus, we permit an initial 'nudging' of the atoms off the high symmetry sites (and see if the quantum mechanical forces tend to restore such high symmetry positions or prefer Jahn-Teller-like displacements). At the same time, one needs to assure that the electronic self-consistency cycle could explore a broader range of wave function symmetries without getting trapped in high-symmetry solutions. To this end we avoid charge density symmetrization during the electronic self-consistent iterations. In the case of the PM phases, one needs to prepare the systems in initial configurations that exhibit unequal $d$ orbital occupations in order to "nudge" the self-consistent solver into an insulating solution. Specifically, we did this by setting[49] equal to one the initial occupation of one of the $t_{2g}$ orbitals in the case of FeO and of two $t_{2g}$ orbitals in the case of CoO. See Appendix B.2 for the details of the nudging protocol. Starting from such an orbital configuration helps the self-consistent solver to converge towards a solution in which the $d$ orbitals mix to form linear combinations whose occupations ultimately are either close to one or zero.

## VI.   The magnetically ordered AFM phases

The AFM phases of MnO, FeO, CoO, and NiO have been studied by DFT[50, 51] as well as its extensions and corrections, including DFT+$U$,[20, 48] hybrid functionals,[52] and SIC.[53] Here we briefly describe our results of the evolution of the band gaps (Fig. 2) and provide the density of states [Fig 3(a,b) and Fig 4(a,b)] to establish a common basis for discussing later the generalized supercells needed to capture the physics of the PM phases. As shown in Fig. 2, in the ideal cubic rock salt structure ($Fm\bar{3}m$ space-group) the crystal field splits the atomic $d$ levels into spin-up and spin-down $t_{2g}$ and $e_g$ levels. The AFM-II magnetic ordering already breaks the cubic space-group magneto-crystalline symmetry even without distortions to the ideal cubic lattice. The lattice relaxations that are experimentally observed in the low-temperature phases of these monoxides[54-56], lower the point-group symmetry of the crystal field at the TM sites with respect to that of the ideal cubic structure. See Appendix C for the details of the relaxed crystal structure of the AFM phases that we obtained by our DFT+$U$ calculations. As a result of the overall symmetry lowering, the $t_{2g}$ and $e_g$ orbitals mix to form linear combinations invariant to the lower-symmetry crystal field. Qualitatively, a similar combined effect of the exchange and crystal field interaction is the mechanism that drives the gap opening in the PM phases and provides a unifying, single-particle description of the insulating character of both the magnetically ordered and magnetically disordered phases. We, therefore, illustrate this mechanism starting with the AFM phases, as well as the hypothetical FM phases, through the same protocol used for the PM phases.

### A.   AFM MnO and NiO.

AFM MnO and NiO exhibit a $R\bar{3}m$ magneto-crystalline structure in which the TM crystal field has the rhombohedral $D_{3d}$ point-group symmetry. A crystal field of this symmetry mixes the

$t_{2g}$ orbitals so as to give the $a_{1g}$ singlet and the $e'_g$ doublet (Fig. 2(c)). The $e_g$ orbitals are symmetry invariant to the $D_{3d}$ symmetry operations and are often indicated as $e''_g$. Figures 3(a)-(b) depict the DOS of MnO and NiO projected onto the cubic $t_{2g}$ and $e_g$ orbitals. In MnO, the $Mn^{2+}$ ions exhibit the $d^5$ electronic configuration which results into the five spin-up $d^\uparrow$ orbitals fully occupied. A band gap opens in MnO between the fully occupied spin-up $d^\uparrow$ orbitals and the empty spin-down $d^\downarrow$ orbitals. In NiO, the $Ni^{2+}$ ions exhibit the $d^8$ configuration which corresponds to the five spin-up $d^\uparrow$ orbitals fully occupied and the $t_{2g}^\downarrow$-derived orbitals also fully occupied. A band gap opens in NiO between the occupied $t_{2g}^\downarrow$-derived levels and the empty $e_g^\downarrow$-derived levels.

### B. AFM FeO and CoO.

$Fe^{2+}$ and $Co^{2+}$ in FeO and CoO are, respectively, in the $d^6$ (meaning one electron in the spin-down $d$ states) and $d^7$ (meaning two electrons in the spin-down d states) configurations. Therefore, without magnetic ordering the degenerate cubic $t_{2g}^\downarrow$ levels in FeO and CoO, respectively, would be partially occupied by one and two electrons making these systems metallic. Our DFT+$U$ calculations, in line with earlier studies[20, 21, 48], show that a gap opens in FeO and CoO already in the undistorted cubic lattice because of the symmetry lowering induced by the AFM-II ordering. FeO opens a gap by occupying the $a_{1g}^\downarrow$ singlet while the $e'_g{}^\downarrow$ doublet is in the conduction. The opposite occurs in CoO with $e'_g{}^\downarrow$ doublet occupied by two electrons while the $a_{1g}^\downarrow$ singlet is in the conduction. FeO[57] and CoO[55, 56] lower their total energies with respect to the ideal cubic lattice through tetragonal distortions of the metal-oxygen coordination octahedra that are accommodated within a monoclinic cell with C2/m space group symmetry (see Fig. A1 and Tab. A1 in the Appendices for a description of the calculated equilibrium crystal structure). In FeO there is a compression of the in-plane bonds and an expansion of the out-of-plane ones, while the opposite occurs in CoO. Figures 4(a)-(b) depict the DOS of monoclinic FeO and CoO projected onto the cubic $t_{2g}$ and $e_g$ orbitals: FeO and CoO continue to be insulating in the stable AFM monoclinic phase as in the higher-energy, AFM undistorted cubic phase. The orbital mixing that occurs because of the tetragonal distortion of the coordination octahedra is reflected by the shape of the spin-down electron density $\rho^\downarrow(\mathbf{r})$. In FeO, $\rho^\downarrow(\mathbf{r})$ has a square-planar shape rotated by 45° degrees around the z axis (see Fig 5(a)). In CoO, $\rho^\downarrow(\mathbf{r})$ has an octahedral shape with the vertical axis lying along the diagonal of the $x$-$y$ plane (see Fig 5(b)). A detailed analysis of $\rho^\downarrow(\mathbf{r})$ in terms of the orbital mixing obtained in the DFT+$U$ solution is performed in Appendix D.1.

### VII. The magnetically disordered paramagnetic phases

The DFT+$U$ calculations of the PM phases of the four monoxides modeled with the magnetic SQS produce insulating solutions with strong magnetic moments at the TM sites (see. Table I for the gaps and magnetic moments at the transition metal sites obtained in these SQS calculations). The minimization of the total energy for the PM phases of NiO, MnO and FeO, modeled by the SQS shows negligible positional atomic displacements relative to to the rock salt positions, thus, no broadening of Bragg diffraction peaks. In the case of CoO, small random displacements (with amplitudes drawn from a normal distribution with a standard deviation of 0.25 Ang.) of the atom positions were needed in the starting supercell to obtain the gapped solution. However, subsequent relaxation decreased the displacements to less than 0.05 Ang. from the ideal rocksalt sites. The projected DOS (PDOS) on the metal $d$ orbitals are depicted in Fig. 3(c,d) for MnO and NiO and Fig. 4(c,d) for FeO and CoO. For the sake of comparing the PDOSs across the whole series of oxides and magnetic phases included in this study, we project the DFT+$U$ wave functions on $t_{2g}$ and $e_g$ orbitals. However, in the polymorphous description which is implemented through the magnetic SQSs, the crystal field at each TM site shows a low point-group symmetry which in turn allows for the $t_{2g}$ orbitals to mix among themselves and possibly also with the $e_g$ orbitals. The magnetic disordered phases of MnO and NiO exhibit gaps that in both cases, as can be seen from the PDOSs in Fig 3(c,d), open between sub-bands that derive predominantly from the $t_{2g}$ and $e_g$ orbitals and are both filled in a similar fashion as in the magnetically ordered AFM phases.

The DFT+$U$ calculations of the magnetic disordered phases of FeO and CoO also gave insulating solutions. The PDOS plots of FeO and CoO in Fig. 4(c) and 4(d), respectively, show that the gap originates in both systems mainly from a splitting in the $t_{2g}$-derived states. An inspection of the density $\rho^\downarrow(\mathbf{r})$ of spin-down electrons of PM FeO and CoO (see Fig. 6(a,b)) and of the eigenvectors of the occupation matrices calculated at the TM sites (see Appendix D.2) shows that the gap is the result of the mixing of the $d$ orbitals in the low-symmetry crystal field that characterizes each site in the disordered phases and of the splitting of these mixed orbitals that lifts the degeneracy of the unperturbed d states. In PM FeO, $\rho^\downarrow(\mathbf{r})$ at the Fe sites (Fig 6(a)) has a square-planar shape lying on one of the Cartesian planes and it is not tilted by 45° around a Cartesian axis as in AFM FeO (see Fig. 5(a) and 5(c) for a comparison between $\rho^\downarrow(\mathbf{r})$ in AFM FeO and PM FeO), with the specific plane varying randomly from site to site. Sites 4 and 14 in Fig. 6(a) are examples of two distinct orientations of $\rho^\downarrow(\mathbf{r})$ around the Fe sites in PM FeO. The analysis of the eigenvectors of the occupation matrices in Appendix D.2 shows that this square planar shape originate from one $t_{2g}$ orbital almost completely filled.

In PM CoO, we observe (see Fig. 6(b)) that $\rho^\downarrow(\mathbf{r})$ shows two types of shapes at the Co sites: a cylinder-like shape (e. g., at site 14 in Fig. 6(b)) aligned to one of the Cartesian axes, and an octahedron-like shape (e.g., at site 4 in Fig. 6(b)), with the apex of the octahedron that, as opposed to the similar octahedral shape that $\rho^\downarrow(\mathbf{r})$ displays in AFM CoO, is tilted away from the Cartesian planes and points into a direction that varies randomly from site to site. Figs. 5(b) and 5(d) display $\rho^\downarrow(\mathbf{r})$ at the Co sites in AFM CoO and PM CoO respectively and show the different orientation of the octahedral shape in the two cases. From an inspection of the eigenvalues of the occupation matrices at the Co sites within the SQS (see Appendix D.2) we observe that at each site two eigenvectors $\phi_{Co}^{(1)}$ and $\phi_{Co}^{(2)}$ have occupations of ~0.9, that is, are almost completely full and provide the dominant contribution to $\rho^\downarrow(\mathbf{r})$ at the Co sites. The two different types of the shape of $\rho^\downarrow(\mathbf{r})$, found for example at site 4 and 14 in Fig. 6(b), originate from two distinct modes of mixing of the $d$ orbitals to which correspond two distinct pairs of (nearly fully) occupied eigenvectors $\{\phi_{Co}^{(1)}, \phi_{Co}^{(2)}\}$. At the sites where $\rho^\downarrow(\mathbf{r})$ shows a cylindrical shape (e.g., site 14 in Fig. 6(b)) the eigenvectors $\{\phi_{Co}^{(1)}, \phi_{Co}^{(2)}\}$ are two of the three $t_{2g}$ orbitals. At sites where $\rho^\downarrow(\mathbf{r})$ shows a tilted octahedral shape, the eigenvectors $\{\phi_{Co}^{(1)}, \phi_{Co}^{(2)}\}$ are normalized linear combinations of the three $t_{2g}$ orbitals with contributions also from the $e_g$ orbitals. Analytic models of the probability densities $|\phi_{Co}^{(1)}|^2$ and $|\phi_{Co}^{(2)}|^2$ and of their sum $|\phi_{Co}^{(1)}|^2 + |\phi_{Co}^{(2)}|^2$ are displayed in Fig. A3 in the Appendices; these models are consistent with the tilted octahedral and the cylindrical shape of $\rho^\downarrow(\mathbf{r})$ at the Co sites in PM CoO obtained in the DFT+$U$ solution.

The modality of orbital mixing and level splitting revealed by the present calculations are analogous in the AFM and PM phase of the respective oxides. This helps explains the fact that the magnetic moments in the SQS PM configurations converge to values whose average is within to less than 1 % of the AFM values (Table I). At the same time, the magnetic disorder decreases the band gap of MnO, NiO, and CoO with respect to the value in the AFM phase.

## VIII. Discussion

### A. The physical picture of gapping of the PM phase

The underlying picture emerging from this work is that if one does not force the macroscopically averaged symmetry on the individual random PM configurations and instead permits larger cells, each of the chemically identical atoms could have its own local 'particle-in-a-box' potential that creates different local charge density distributions and magnetic moments. The application of DFT+$U$ to such polymorphous description of the PM phases shows that the exchange interaction and the low-symmetry crystal field make the orbitals mix and split in ways that fully resolve the degeneracy of the unperturbed spherical ion. The application of the DFT+$U$ method with finite $U$ is needed to correct the self-interaction error. Therefore, one is not at liberty to dial the "$U$" parameter down to U=0 because this means reverting to plain vanilla LDA or GGA and, therefore, losing altogether the self-interaction correction effect that is delivered by the "+$U$" part of the DFT+$U$ functional when $U$ is different than zero. The exchange and crystal field interactions are local effects that do not need a long-range magnetic ordering to mix and split the d levels, and as such these local effects drive the opening of a gap in the overall magnetically disordered phases. In a disordered phase one must allow for identical chemical atoms to have the opportunity to experience different local structural environments, or else magnetic moments have no chance to develop. The SQS captures disorder and is polymorphous and works for all four cases tried. The FM phase is ordered and therefore not relevant to the description of the PM phase. The band splitting that is the result of exchange and crystal field effects produces split-off bands that are single-particle states and should not be confused with the lower and upper Hubbard bands[58, 59] that characterize the solutions of the Hubbard model. The Mott mechanism requires that the electrons move across the lattice forming states on certain atomic sites with doubly occupied $d$ orbitals and empty $d$ orbitals on other sites while the overall charge number is conserved. These types of excited configurations correspond respectively to the upper and lower Hubbard bands, which are truly "dynamic" charge bands and correspond to many-body configurations, whereas our band (i.e., single particle) description does not produce this description.

## B. Comparison with other approaches

In the monomorphous description underlying the DMFT[16], one gets a correlation-induced magnetic moment on each of the symmetry-equivalent TM sites in the NaCl unit cell, with the moment fluctuating in time to give the zero moment PM phase. Within correlated theories such as DMFT, the gap of the AFM and PM phases of these oxides emerges because the $d$ electrons become localized on atomic orbitals (not in band states) even in the absence of spatial symmetry breaking. In band theory, on the other hand, a monomorphous description gives incorrectly zero moments and zero gaps statically. This led to the well-known textbook expectation that the phenomena of gapping in the PM phase forces upon us a departure from band theory. However, we point out that a proper polymorphous configurational description gives a gap in the PM phase already at the level of a generalized Kohn-Sham theory. The calculation of macroscopic observables, e.g., the band gap, of the PM phases requires performing averages over the appropriate statistical ensemble. It is wrong to substitute the ensemble average $\langle P \rangle = \sum_n c_n P(\sigma_i^{(n)})$, where the summation defined over the full configurational space of the system, with the property $P(\Sigma_0)$ of the averaged structure $\Sigma_0$, which in the case of the monoxides studied here is the non-magnetic NaCl structure. We conclude that dynamic correlation is not forced upon us just by the need to describe gapping in the PM phase. The demonstration that going past the standard average non-magnetic structure of the PM phases by using a model that captures the realizations of the magnetic configurations is significant because of its implication on establishing the minimal theoretical formalism needed to capture the formation of strong local moments and the insulating character of MnO, NiO, CoO, and FeO.

Previously the paramagnetic phases, including those of the present TM oxides, have been modeled by the disordered local moments (DLMs)[60-62] approach which has been implemented within the single-site coherent-potential approximation (CPA).[11, 26, 63, 64] This approach assumes that the Schrodinger equation potential seen by chemically equivalent atoms in the disordered PM phase are all equal even though such atoms have distinctly different local environments (such as number of neighbors with spin up vs. spin down, see Fig. 1). This picture automatically ignores the existence of inhomogeneous distribution of moments and charges. This is valid only when the local environment flips its spin so fast that a central atom does not distinguish if its environment is made of up spins or down spins but all can be described as some average. This is unlikely to be the case in insulators (such as Mott insulators) where the screening is ineffective. This DLM view leads to equal local moments on all TM atoms irrespective of their environments. The DLM approach is virtually equivalent to the so-called "Hubbard III" approximation[14] in regard to the treatment of the spin disorder. DLM and Hubbard III are in turn related to the DMFT approach, with the difference that DLM and Hubbard III are unable to describe quantum fluctuations which are instead described by DMFT[16]. Similarly, to the DLM description, DMFT is inherently a single-site theory in which all sites of a given species (e.g., all the Co sites in paramagnetic CoO) are geometrically equivalent.

The Special Quasi-Random Structure construct is an effective way to establish a physically grounded representation of the random magnetic configuration for three reasons. First, an SQS is constructed so that a property calculated using it is a close estimate of the ensemble statistical average that would be required to calculate that property for a fully disordered phase. Therefore, using one SQS one can obtain reliable estimates of ensemble averages, i.e., the relevant quantities for the paramagnetic phases, by calculating one configuration instead of many randomly-generated configurations. Second, an SQS allows for a variety of local magnetic environments and for multiple patterns of uneven $d$ orbital occupations that both concur to breaking the cubic symmetry and fully lifting the degeneracy of the $d$ orbitals. Finally, it is straightforward to construct SQS that represent the property of imperfectly-random ensembles, i.e., those that have short range order (SRO) and are thus better representative of PM phases closer to the Néel temperature. Instead of constructing the SQS by fitting to the analytically known random pair and many body correlation functions (see Appendix A), one can fit to independently measures or calculated correlation functions that incorporate SRO.[65]

In conclusion, in the present study we find that the DFT+$U$ method, which is a generalized Kohn-Sham approach reproduces the insulating character and on-site magnetic moments of the prototypical Mott insulators MnO, NiO, CoO, and FeO when applied to SQSs, which approximate closely the ensemble average over the random magnetic configurations.

**Acknowledgements**

The work of A. Z. was supported by Department of Energy, Office of Science, Basic Energy Science, MSE division under Grant No. DE-FG02-13ER46959 to CU Boulder. He would like to thank Chris Marianetti, David Singh, Sohrab Ismail Beigi, Gabi Kotliar, and Olle Helleman for very interesting and extremely useful discussions. G. T. would like to thank also Prof. J. Ketterson of Northwestern University for useful discussions. The crystal structure figures displayed in this article were generated using the software VESTA.[66] In this research, G.T. used resources of the National Energy Research Scientific Computing Center (NERSC), a U.S. Department of Energy Office of Science User Facility, supported by the Office of Science of the U.S. Department of Energy under Contract No. DE-AC02-05CH11231.

|  |  | $R\bar{3}m$ | | | | C2/m | | | |
| --- | --- | --- | --- | --- | --- | --- | --- | --- | --- |
|  |  | MnO | | NiO | | FeO | | CoO | |
|  |  | Exp. | DFT+$U$ | Exp. | DFT+$U$ | Exp. | DFT+$U$ | Exp. | DFT+$U$ |
| AFM | μ (μ$_B$) | 4.58 | 4.64 | 1.9 | 1.68 | 4.0 | 3.71 | 3.8-3.98 | 2.72 |
|  | E$_{gap}$ (eV) | 3.5 | 1.88 | 3.5 | 3.00 | 2.1 | 1.66 | 2.8 | 2.63 |
| FM | μ (μ$_B$) | - | 4.68 | - | 1.75 | - | 3.76 | - | 2.78 |
|  | E$_{gap}$ (eV) | - | 0.82 | - | 2.56 | - | 1.36 | - | 2.06 |
| PM | μ (μ$_B$) | - | 4.65 | - | 1.70 | - | 3.73 | - | 2.75 |
|  | E$_{gap}$ (eV) | 3.7 | 1.22 | 4.1 | 2.16 | 2.5 | 1.70 | 2.4 | 2.25 |

**Table 1**. Experimental and DFT+$U$ calculated band gaps E$_{gap}$ and magnetic moments of the transition metal atoms of MnO, NiO, FeO, and CoO in the following phases: (i) AFM with rhombohedral $R\bar{3}m$ space-group symmetry for MnO and NiO and monoclinic $C2/m$ space-group symmetry for FeO and CoO), (ii) FM with cubic symmetry calculated using the same volume per formula unit that was found by total energy optimization for the AFM phase, and (iii) the PM phases calculated by a 64-atom 2×2×2 SQS with a volume per formula unit that is equal to the optimal volume per formula unit found for the AFM phases. The following are the sources for the experimental values: Ref.[10] for the band gaps of the AFM phases; Ref.[11] for the band gaps of the PM phases; Ref.[12] for the magnetic moments in the AFM phases. We were unable to find in the literature the experimental values of the magnetic moments of the PM phases.

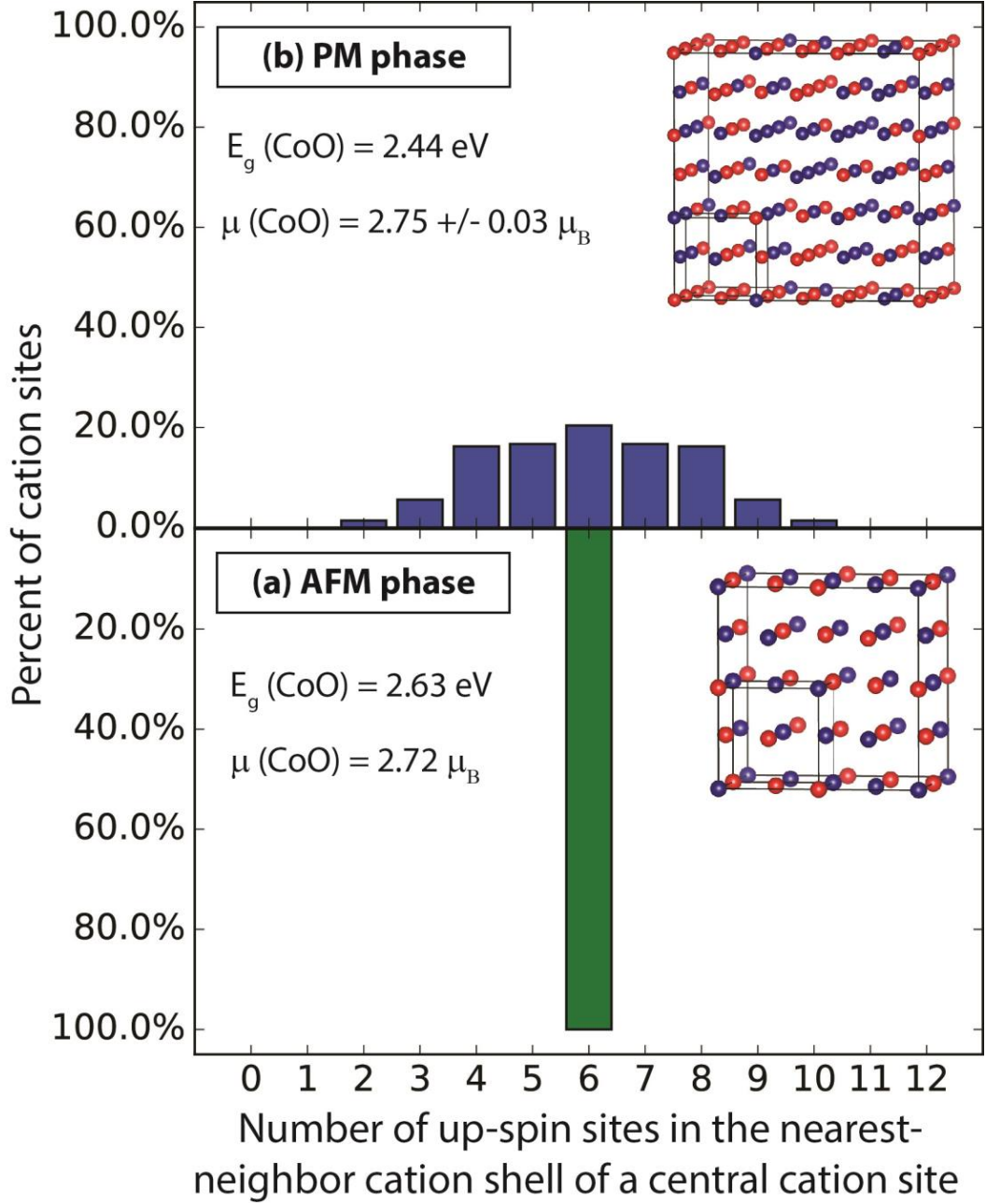

**Figure 1.** Percent fraction $F(n^{\uparrow}_{NN})$ of cation sites as a function of the number $n^{\uparrow}_{NN}$ of spin-up nearest-neighbor (NN) metal site of a central cation site in (a) the AFM-II phase and (b) the PM phase modeled here by the 216-atom rock-salt SQS shown in the insert (in this model only the metal sub-lattice of the underlying rock salt structure is shown). In the AFM phase $n^{\uparrow}_{NN}= 6$ for all cations while in the PM phase $n^{\uparrow}_{NN}$ varies between 0 and 12 and $F(n^{\uparrow}_{NN})$ approximates a binomial distribution. The frequency of local environment types $F(n^{\uparrow}_{NN})$ was calculated averaging over a 216-atom SQS with $A_{0.5}B_{0.5}$ composition and its complementary $B_{0.5}A_{0.5}$ spin configuration. The band gap and the TM magnetic moment are reported for CoO in the AFM phase and the PM phase which was calculated using the displayed 216-atom 3×3×3 SQS.

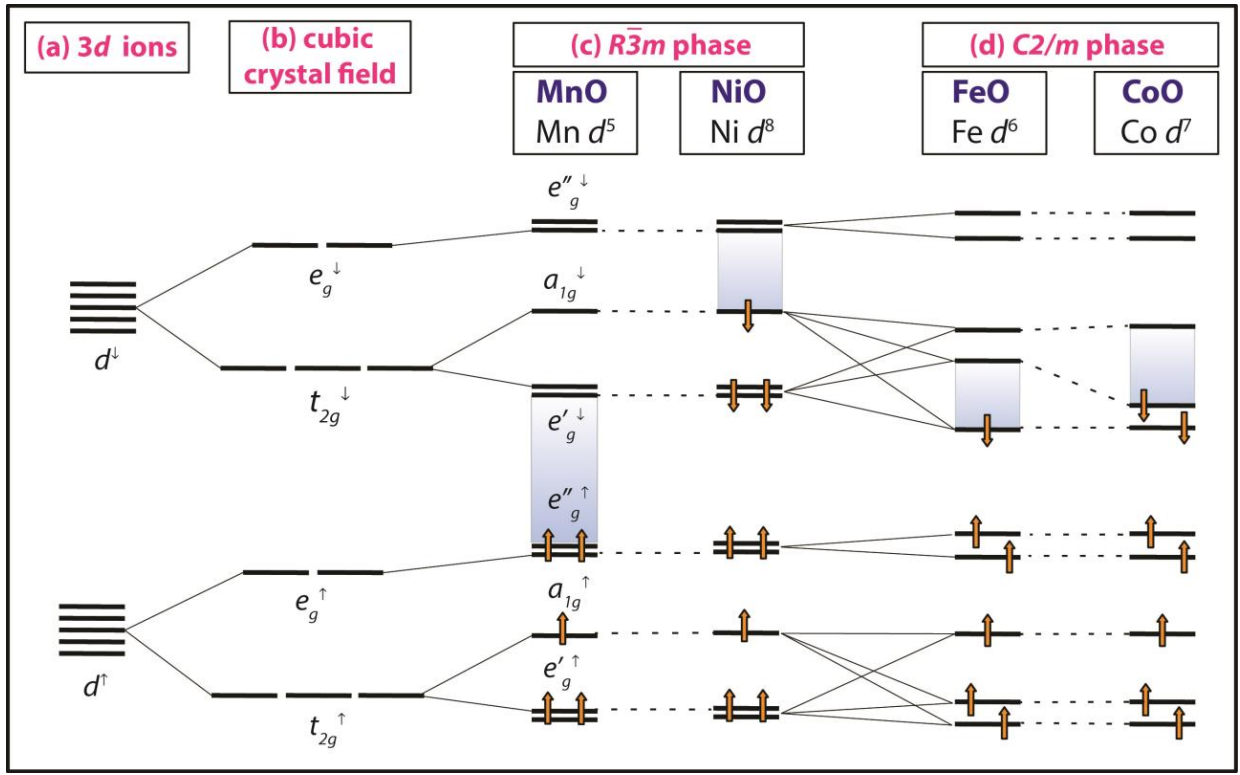

**Figure 2**. Schematic of the sequence of level splittings and combinations for the $d$ orbitals in MnO, NiO, FeO, and CoO as the exchange coupling and the crystal field of the symmetry appropriate to each phase are progressively imposed: (a) Splitting of the $d$ orbitals into the transition-metal atoms subjected to exchange coupling. (b) Splitting of the spin-up and spin-down $d$ levels subjected to a cubic $O_h$ crystal field: this is the case of a rock salt structure with hypothetical $Fm\bar{3}m$ magnetic ordering. (c) Splitting of the $d$ level in the $D_{3d}$ crystal field in the distorted rock-salt lattice with the rhombohedral $R\bar{3}m$ AFM-II magnetic ordering. (d) Splitting of the $d$ orbitals in a tetragonal crystal field as in the monoclinic $C2/m$ phases of FeO and CoO. Note that in this schematic we emphasize the effect of the relevant interaction in progressively removing the degeneracy of the $d$ orbitals, while we do not intend to reproduce to scale the position of the spin-up and spin-down energy levels. The reader should inspect the projected DOSs of Fig. 3 and 4 to obtain the actual calculated energy position of the $d$ bands.

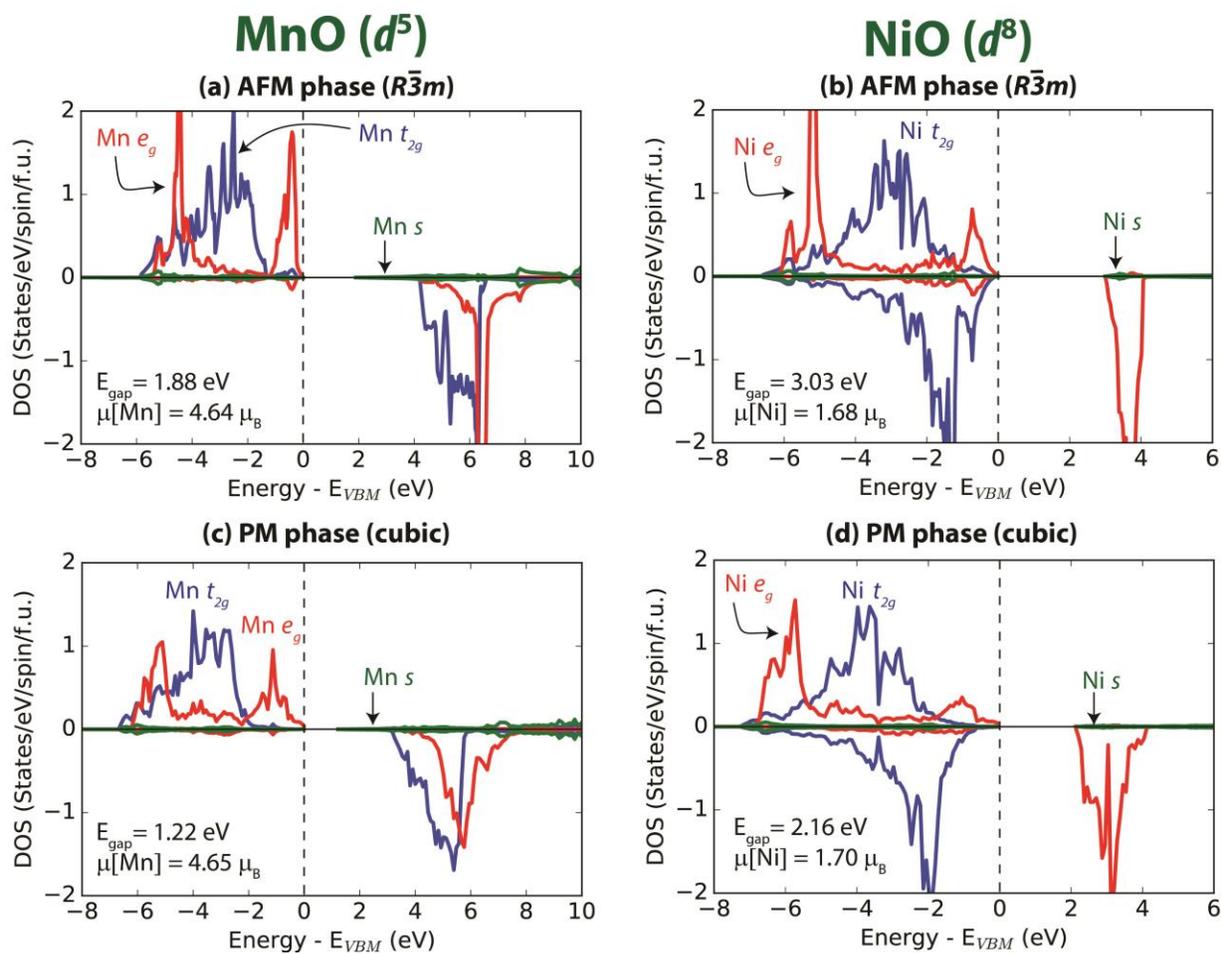

**Figure 3**. Projected density of states (PDOS) on the transition metal $s$ and $d$ orbitals ($t_{2g}$ and $e_g$ components) calculated by DFT+$U$ ($U$=5 eV) for MnO and NiO in (a)-(b) the AFM phase with fully relaxed $R\bar{3}m$ structures, and (c)-(d) the PM phase modeled by a cubic 64-atom 2×2×2 SQS. The lattice parameters of the SQSs are set so that the volume per formula unit is equal to the calculated volume per formula unit of the DFT+$U$ relaxed $R\bar{3}m$ structures of MnO and NiO.

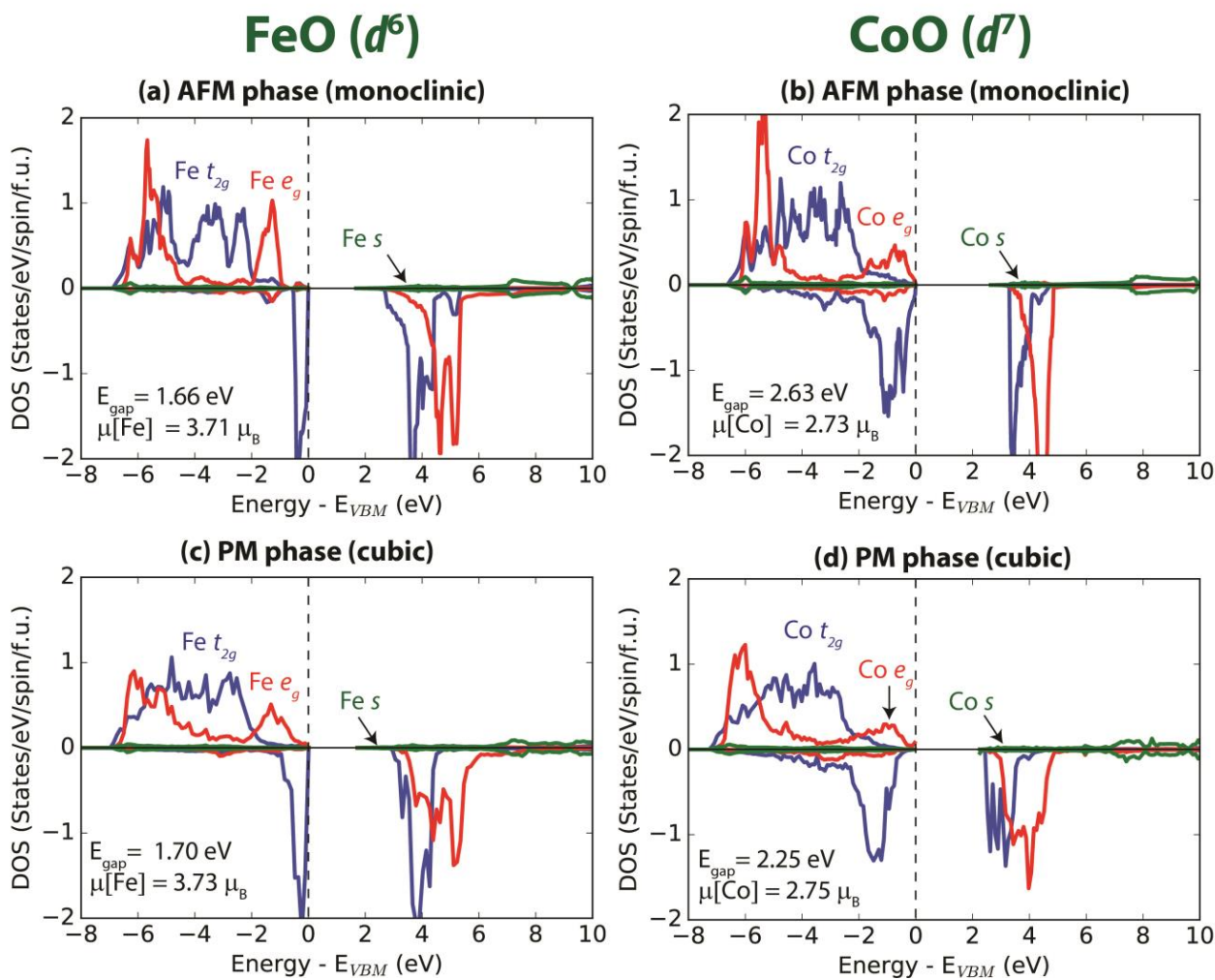

**Figure 4**. Projected density of states (PDOS) on the transition metal $s$ and $d$ orbitals ($t_{2g}$ and $e_g$ components) calculated by DFT+$U$ ($U$=5 eV) for FeO and CoO in (a)-(b) the AFM phases with the fully relaxed monoclinic $C2/m$ structures, and (c)-(d) the paramagnetic phases modeled by a cubic 64-atom 2×2×2 SQS. The lattice parameters of the SQSs are set so that the volume per formula unit is equal to the calculated volume per formula unit of the DFT+$U$ relaxed $C2/m$ structures of FeO and CoO.

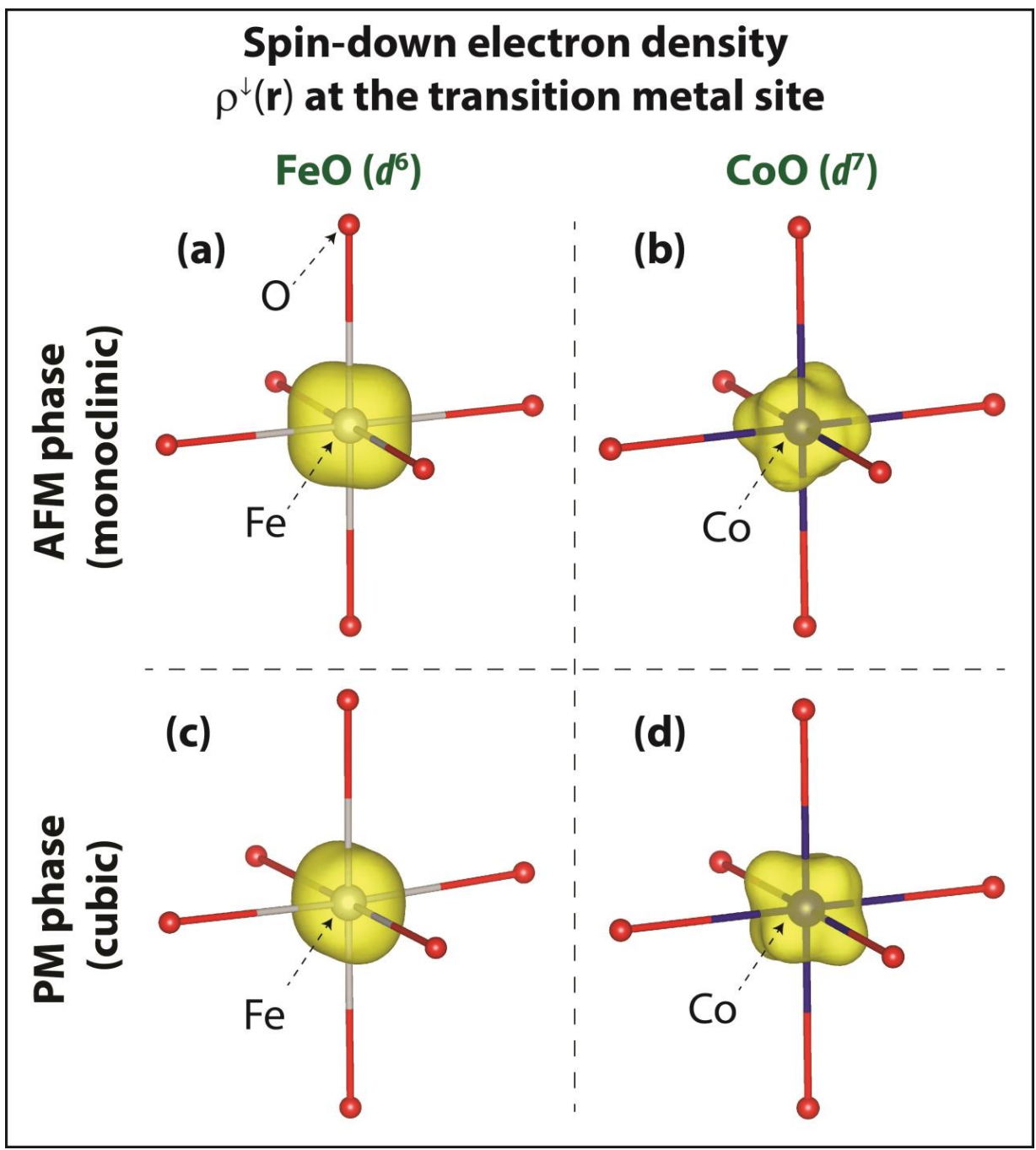

**Figure 5.** Panel (a) and (b) depict the minority spin electron density $\rho^\downarrow(r)$ at the Fe and Co site in the AFM monoclinic phases of, respectively, FeO and CoO calculated by DFT+$U$. Panel (c) and (d) depict the minority spin electron density $\rho^\downarrow(r)$ at the Fe and Co site in the PM phase of, respectively, FeO and CoO modeled by the magnetic 64-atom cubic SQS used for the calculation of the PDOS shown in Fig. 5(c,d). The full $\rho^\downarrow(r)$ of PM FeO and CoO in the SQS is shown in Fig. 6(a,b).

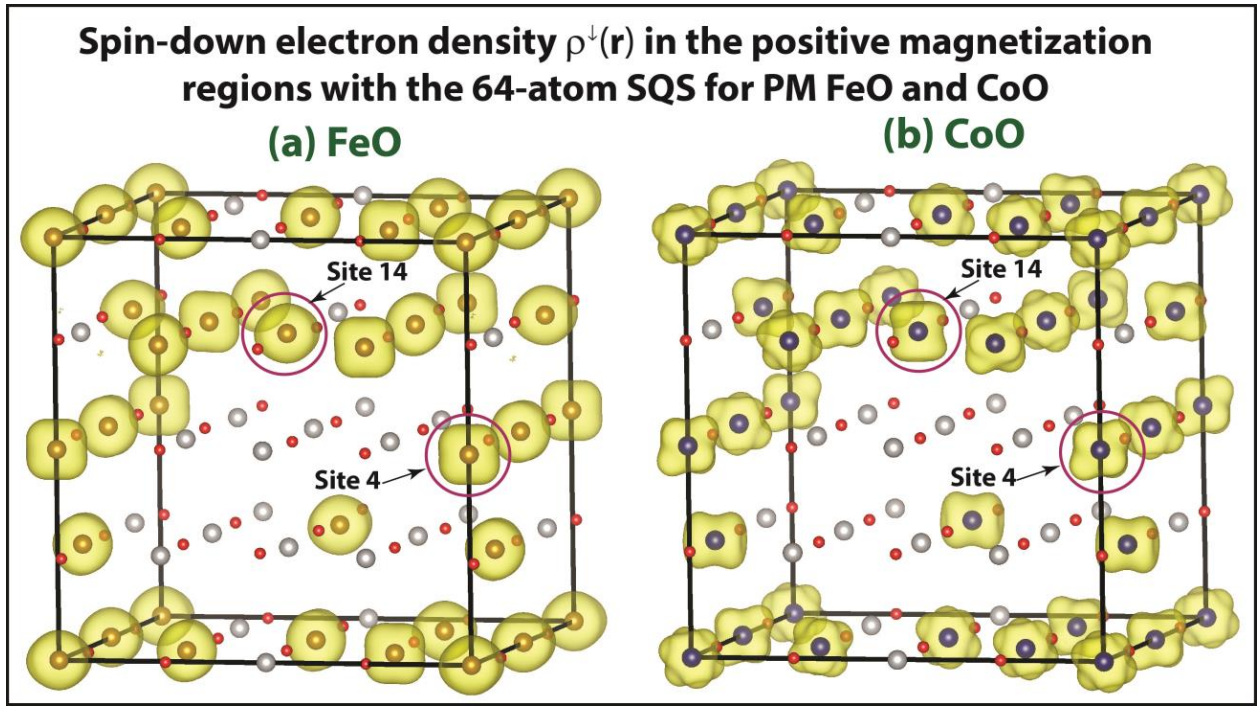

**Figure 6**. Minority spin electron density $\rho^\downarrow(r)$ in the regions with positive magnetization $m(r) = \rho^\uparrow(r) - \rho^\downarrow(r) > 0$ within the magnetic 64-atom SQS cell used in the DFT+$U$ (U=5 eV) calculations of the PM phases of (a) FeO and (b) CoO. To avoid visual clutter we masked out the regions of space within spheres of 1 Å of radius centered at the oxygen sites.